\documentclass[journal,twoside,onecolumn]{IEEEtran}

\usepackage{cite}
\usepackage{authblk}
\usepackage{setspace}
\usepackage{physics}
\usepackage{tikz}
\usetikzlibrary{quantikz2}
\usepackage{amssymb}
\usepackage{amsmath}
\usepackage{booktabs}
\usepackage{comment}

\title{A Study on Quantum Graph Neural Networks Applied to Molecular Physics}
\author[1]{Simone Piperno}
\author[1]{Andrea Ceschini}
\author[2,3]{Su Yeon Chang}
\author[2]{Michele Grossi}
\author[2]{Sofia Vallecorsa}
\author[1]{Massimo Panella}

\affil[1]{Department of Information Engineering, Electronics and Telecommunications, University of Rome ``La Sapienza'', Rome, Italy}
\affil[2]{European Organization for Nuclear Research (CERN), Geneva, Switzerland}
\affil[3]{Institute of Physics, Ecole Polytechnique F\'ed\'erale de Lausanne (EPFL), Lausanne, Switzerland}

\begin{document}

\maketitle

\begin{abstract}
This paper introduces a novel architecture for Quantum Graph Neural Networks, which is significantly different from previous approaches found in the literature. The proposed approach produces similar outcomes with respect to previous models but with fewer parameters, resulting in an extremely interpretable architecture rooted in the underlying physics of the problem. The architectural novelties arise from three pivotal aspects. Firstly, we employ an embedding updating method that is analogous to classical Graph Neural Networks, therefore bridging the classical-quantum gap. Secondly, each layer is devoted to capturing interactions of distinct orders, aligning with the physical properties of the system. Lastly, we harness SWAP gates to emulate the problem's inherent symmetry, a novel strategy not found currently in the literature. The obtained results in the considered experiments are encouraging to lay the foundation for continued research in this field.
\end{abstract}

\section{Introduction}
\label{chap:1} 
Until recently, traditional computers have played a key role in advancing our understanding of molecular structures \cite{mgcn, Wu_gnn, reiser2022graph, Choudhary_2022}. However, as molecules increase in size, computational chemistry faces challenges in accurately predicting molecular properties \cite{mcardle2020quantum}. Quantum computers, leveraging qubits manipulation and quantum properties such as superposition and entanglement, offer a novel computational approach, with the potential to significantly accelerate calculations for previously intractable problems \cite{cao2019quantum, shang2023towards}.

In the last few years, the convergence of quantum computing and Machine Learning (ML) has given rise to Quantum Machine Learning (QML) \cite{qml}, a groundbreaking approach that utilizes quantum mechanics to address complex problems across multiple scientific disciplines. QML combines quantum computing with ML, employing quantum algorithms to enhance model training and execution \cite{qml}. This synergy has the potential to transform fields like drug discovery \cite{batra2021quantum}, materials science \cite{sauceda2022bigdml}, and molecular physics \cite{huang2020quantum} by efficiently exploring complex molecular landscapes.

Within this context, Quantum Neural Networks (QNNs) have emerged as a promising trend in QML \cite{chen2020novel,ceschini_2021}. By projecting data into complex high-dimensional Hilbert spaces, leveraging qubits' superposition and entanglement, QNNs efficiently handle large-scale heterogeneous data and conduct high-dimensional processing, resulting in faster computation and lower error rates \cite{tacchino2019artificial,abbas2021power}. 
Quantum technologies for QNNs range from superconducting qubits \cite{tacchino2020quantum} to photonic systems \cite{wanQuantumGeneralisationFeedforward2017} and trapped-ion devices \cite{huber2021realization}.
Since fault-tolerant quantum computers are still years away, researchers made significant advancements in the design of quantum algorithms meant to be executed on Noisy Intermediate-Scale Quantum (NISQ) devices \cite{lau2022nisq}. Variational Quantum Algorithms (VQAs) have thus emerged as a clever approach to exploit these near-future quantum systems \cite{cerezo2021variational}. VQAs utilize hybrid quantum-classical optimization techniques, employing shallow quantum circuits that are less susceptible to noise and decoherence. These algorithms exhibit promising potential across various domains such as quantum chemistry, machine learning, and optimization tasks \cite{singh2023benchmarking,adebayo2023variational,ceschini2022hybrid,moll2018quantum}.

Our paper leverages QML, specifically VQAs, to efficiently predict intramolecular forces and total molecular energy in water molecules ($\mathrm{H_2 O}$).
In particular, a Quantum Graph Neural Network (QGNN) is appropriately designed to reflect the problem's geometry.
Our model surpasses other comparable quantum and classical models in interpretability: a fundamental strength of our architecture lies in its adeptness at exploiting the inherent symmetry of the problem. This is achieved through the strategic utilization of SWAP gates between the two hydrogen atoms, effectively enforcing a form of permutation invariance driven by the weights.
Our approach allows us to attain performances on par with state-of-the-art methods while utilizing approximately $\frac{3}{5}$ of their parameters for intramolecular forces prediction. 
However, due to error propagation from force calculations, our approach yields energy prediction performances an order of magnitude superior to those achieved by current leading techniques.

Although the water molecule serves as a case study for this paper, due to its fundamental role in chemistry and suitability for quantum simulations, we envision that the principles and methodologies developed herein can be readily extended to explore more complex molecular systems with equal efficacy. This may not only enhance our comprehension of chemical processes, but also open avenues for innovative applications in fields ranging from materials science to computational chemistry.

The main contributions of this paper are:
\begin{itemize}
    \item to design and simulate a novel QGNN architecture, capable of efficiently modeling molecular properties;
    \item to offer valuable insights into the interpretability of our QGNN model, shedding light on its inner workings;
    \item to contribute significantly to comprehend the potential impact of QML and QGNNs in the field of computational chemistry.
\end{itemize}

The rest of the paper is organized as follows.
Sect.~\ref{chap: Chapter 2} reviews relevant literature, highlighting the current state-of-the-art in classical ML and QML chemistry applications.
Sect.~\ref{chap: Chapter 3} covers dataset preparation and augmentation techniques.
The QGNN architecture is meticolously detailed in Sect.~\ref{chap: Chapter 4}, while Sect.~\ref{chap: Chapter 5} focuses on experimental assessment, covering setup, training, evaluation metrics, and results.
Finally, conclusions and future avenues are drawn in Sect.~\ref{chap: Chapter 6}.
    
\section{Related Works}
\label{chap: Chapter 2} 

Graph Neural Networks (GNNs), which were first introduced in \cite{mp_gnn}, are an emerging class of ML models that are well-known for their ability to handle graph-structured data. Their applications span from social networks and molecular structures to recommendation systems \cite{wu2020comprehensive}.
In the context of chemistry, ML tasks involve graph-wide properties regression or graph classification, prevalently. 
When the graph's 3D structure is provided, inherent symmetries like rotational or translational symmetries may be relevant to the problem. Incorporating these inductive biases into the ML model stands as a crucial determinant of its performance \cite{atz2021geometric}.

The message-passing neural network (MPNN) framework \cite{chem_mpnn} aims at updating nodes' embeddings of a given graph based on the neighbours' embeddings of each node.
For instance, \cite{chem_mpnn} used an MPNN with gated graph-neural-network (GGNN)-type message passing \cite{li2015gated}, while \cite{yang2019analyzing} introduced a variant of the model called directed MPNN, which uses message passing of directional link features. Authors in \cite{graph_track} present a novel GNN-inspired framework for particle track reconstruction, which employs a classical architecture where node and edge networks are alternated.
Initially, data preprocessing occurs through Fully Connected layers, forming the input network. Subsequently, the graph model undergoes training by iteratively alternating between the edge and node networks. These networks serve to update edge and node weights respectively, via Fully Connected layers as well. 
Finally, a final edge network is applied, which is crucial for predicting the presence or absence of edges, given the task at hand.
Such a GNN-inspired framework seems to be particularly suited for space-point data prediction, providing satisfactory results with a flexible architecture.

In the case of molecules, it is intuitive to represent them as graphs, where atoms are nodes and bonds are represented by links. Consequently, ML models can in principle predict certain molecules' quantum features through the three-dimensional positional data of individual atoms. 
For example, SchNet \cite{p20} takes as inputs the three-dimensional positions and nuclear charges of every atom in the given molecule, and the distances between atoms serve as filters that are used to update the atomic features. Notably, the network generating these filters is trainable.

From a quantum viewpoint, the exploration of quantum algorithms suitable for NISQ systems has led to the emergence of VQAs, whose functioning is based on a hybrid quantum-classical optimization framework \cite{vqa,qcl} centered around Variational Quantum Circuits (VQCs), also known as Parametrized Quantum Circuits (PQCs). The latters are considered promising contenders for unlocking the potential of quantum computing and gaining a quantum advantage \cite{Di_Marcantonio_2023, Liu_2021}. VQCs employ a hybrid quantum-classical approach, iteratively refining a parameterized quantum circuit with the aid of a classical co-processor. This method facilitates the efficient design and implementation of shallow quantum circuits on existing NISQ devices, alleviating the impact of quantum noise due to the restricted number of quantum gates in VQCs.
Diverse circuit architectures and ansatz variations have been suggested for VQCs \cite{Benedetti_2021, farhi2014quantum, Patil_2022}, demonstrating impressive performance and resilience to noise across various optimization tasks and real-world applications \cite{adebayo2023variational}.
In \cite{Schuld_2020}, a VQC-based circuit-centric quantum classifier proficiently classified quantum-encoded data and demonstrated resilience against noise. Another potential application is presented in \cite{qcl}, where a VQC effectively approximated high-dimensional regression and classification functions with a constrained number of qubits.

VQCs are well-suited for realizing QNNs with constraints on qubit count. The encoding strategy and the selection of the circuit ansatz are critical for achieving superior performance compared to classical Neural Networks (NNs).  Instances of QNNs, like those presented in \cite{macalso_qnn,zhao2020qdnn}, have demonstrated heightened representation capacity compared to classical deep NNs. 
Despite their potential, QNNs encounter challenges impacting their performance and constraining their impact. Among these difficulties there is the vulnerability to quantum noise, which gets worse as circuit depth increases \cite{wang2023quantumnat,liang2021noise}, the occurrence of barren plateaus depending on the ansatz and qubit count, diminishing model trainability \cite{McClean_2018, holmes_expr, Cerezo_2021}, and limitations in data encoding on NISQ devices, especially when handling a substantial number of features.

Regarding the field of molecular properties forecasting, researchers in \cite{oriel} devised a QNN architecture and applied it across a spectrum of molecules, including $\mathrm{LiH}$, $\mathrm{H_2O}$, and $\mathrm{H_3 O^+}$, aiming at forecasting their force fields. 
They showed that the quantum models exhibit larger effective dimension with respect to classical counterparts and can reach competitive performances, thus pointing towards potential quantum advantages in natural science applications via QML.

On the other hand, very few works approached to the realm of QGNNs. Authors in \cite{1} introduced a QGNN characterized by Hamiltonian operators responsible for generating rotations that align with the topology of the problem graph.
These models were used for various toy problems that included Hamiltonian dynamics learning and graph clustering.
\cite{2} pioneered the concept of graph-structured quantum data, where individual nodes are associated with quantum states, and links are established between two quantum states when they are within a specified information-theoretical distance of each other. This research contributed to novel designs for loss functions and training methodologies, particularly focused on leveraging dissipative QNNs.
Researchers in \cite{3} integrated state preparation, quantum graph convolution, quantum pooling, and quantum measurements into a VQC. 
This framework was applied to the classification of handwritten digits, showcasing its applicability in real-world tasks.
Until now, existing approaches have overlooked the scalability of quantum circuits with respect to the underlying problem. For this reason, a hybrid classical-quantum GNN was introduced in \cite{4}. Such an approach involves updating the link and node features of an input graph using link and node networks, similar to classical GNNs presented in \cite{graph_track}. However, what sets this model apart is the integration of a QNN layer between classical Fully Connected layers to create the link and node networks. This hybrid model was put to the test in particle track reconstruction at the Large Hadron Collider, and it demonstrated performance results comparable to classical models. Finally, a variation of the approach outlined in \cite{1} harnessed the principle of equivariance from geometric ML, culminating in the creation of Equivariant Quantum Graph Circuits (EQGCs) \cite{mernyei2022equivariantquantumgraphcircuits}. The latters are designed to maintain equivariance with respect to permutations, allowing for the efficient handling of graph data.

Our proposed QGNN architecture, which combines interpretability and performance with a foundation in the underlying physics, marks a substantial divergence from previous methods. The innovative methodology outlined in this paper holds great promise for handling complex graph data while maintaining interpretability. By harnessing SWAP gates to mimic symmetries and introducing a groundbreaking embedding update technique, this approach offers hope for a variety of applications in the QGNNs' realm.

\section{A Case Study on Molecular Physics}
\label{chap: Chapter 3}
\begin{figure}[!ht]
    \centering
    \includegraphics[width=0.7\textwidth]{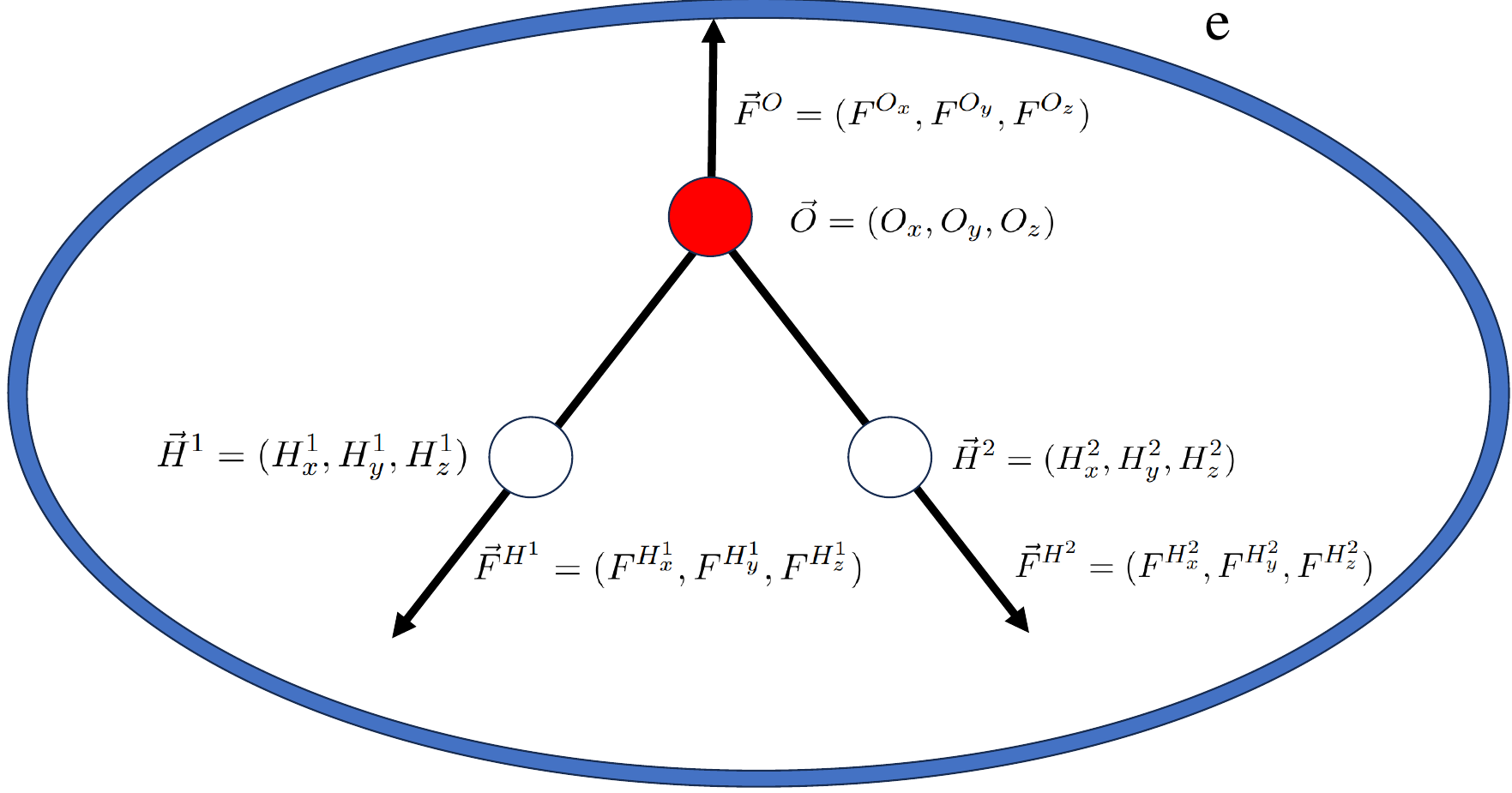}   
    \caption{Graphical representation of a single dataset entry. The latter describes the arrangement of atoms within an $H_2O$ molecule, showcasing the coordinates and forces acting on each atom.}
    \label{fig:dataset}
\end{figure}

This study focuses on leveraging a QGNN in order to predict intramolecular forces and energy within a $H_2O$ molecule.
In particular, the same dataset as the one introduced in \cite{oriel} is employed for the experiments. This dataset comprises $n=1000$ individual entries, each representing a distinct $H_2O$ molecule.
The features, i.e. the coordinates of the atoms in the molecule, are referred to as $\{\alpha^i_c\},$ where $\alpha \in \mathcal{A}=\{O,H^1,H^2\}$ represents the type of atom, $c \in \mathcal{C}=\{x,y,z\}$ is the coordinate axis and $i$ represents the $i$-th datapoint. As this is a supervised learning task, 10 target variables associated to the previously defined features are present: one for the energy, denoted as $e_i$, which represents a global property of the $i$-th molecule, and nine force components. These force components are the projections of resultant forces onto each atom $\vec{F}_i^{\alpha}$ along the $(x,y,z)$ axes, represented as $F_i^{\alpha_c}$. A graphical illustration of a single sample from the dataset is depicted in Fig.~\ref{fig:dataset}.

Given this configuration, every data point presents a unique challenge in the form of 9 input coordinates and 10 target variables for prediction. This setup transforms the task into a regression problem, with the primary objective being the prediction of both the force components and the energy associated with each molecule.

\subsection{Data Preprocessing}
In preparation for the QML task, it was essential to preprocess the dataset properly. This preprocessing step aimed to ensure compatibility with the quantum circuit and quantum measurements while preserving the essential information contained within the data.
In order to make input features suitable for quantum data encoding, feature rescaling was performed. Specifically, the input features were scaled to the range $[0, 1]$, which made them appropriate for a correct and interpretable encoding in the quantum circuit.
In addition to feature rescaling, also the target variables in the dataset were addressed: the outputs were normalized by constraining both the force components and the energy values to the range $[-1, 1]$. This rescaling allowed to represent the target variables as expected values with respect to a given observable following qubits measurement.
The last step of data preprocessing consisted of building a tensor of distances $\textbf{D} \in \mathbb{R}^{n \times 3 \times 3} $ which is needed in order to encode in the circuit the information about distances between atoms. For each data point, the tensor $\textbf{D}$ encapsulates a $3 \times 3$ matrix that conveys atom-to-atom distance information. A representative entry within the $\textbf{D}$ tensor is exemplified in Table~\ref{Tensor D}, where $\textbf{D}$ tensor's entries are rescaled in $[0,1]$ range as well.
\begin{table}[!ht]
\centering
\caption{An Example of Entries for Tensor \textbf{D}}
\renewcommand{\arraystretch}{1.4}
\begin{tabular}{|c|c|c|}
\hline
$\abs{O_x-H^1_x}$ & $\abs{O_x-H^2_x}$ & $\abs{H^1_x - H^2_x}$ \\
\hline
$\abs{O_y-H^1_y}$ & $\abs{O_y-H^2_y}$ & $\abs{H^1_y - H^2_y}$ \\
\hline
$\abs{O_z-H^1_z}$ & $\abs{O_z-H^2_z}$ & $\abs{H^1_z - H^2_z}$ \\
\hline
\end{tabular}
\label{Tensor D}
\end{table}

\subsection{Data Augmentation}
Before splitting the dataset into training, validation, and test sets, the dataset was augmented for better model training. 
Each datapoint is expected to be rotational invariant: by rotating the molecule, the resulting forces on each atom are rotating along with it while the total energy stays invariant. This means that inputs and labels can be rotated in the same way in order to generate extra datapoints.
This is easily done by making use of Rodrigues' rotation formula, which is a mathematical method for expressing a rotated vector in the three-dimensional space. It provides a way to compute the resulting vector after applying a rotation to an initial vector and can be expressed as follows. Suppose to have an initial 3D vector $\vec{v} = (x, y, z)$, to compute the rotated vector $\vec{v'}$, which is the result of a rotation of $\vec{v}$ by an angle $\theta$ around an arbitrary unit vector $\vec{k} = (a, b, c)$, one can use the Rodrigues' rotation formula, which is defined as:
\begin{equation}
   \vec{v'} = \vec{v} \cos{\theta} + (\vec{k} \times \vec{v}) \sin{\theta} + \vec{k}(\vec{k} \cdot \vec{v})(1-\cos{\theta}).
\end{equation}

Applying the Rodrigues' formula $9000$ datapoints were generated thus passing from a dataset comprised of $n=1000$ entries to a dataset having $n'=10000$ entries.
The such obtained dataset was then split into train, validation and test sets, keeping $80\%$ of the data for training, $10\%$ for validation and the remaining $10\%$ for testing.

\section{Proposed Methodology}
\label{chap: Chapter 4}

\subsection{QGNN}
\begin{figure}[!ht]
    \centering
    \includegraphics[width=0.9\textwidth]{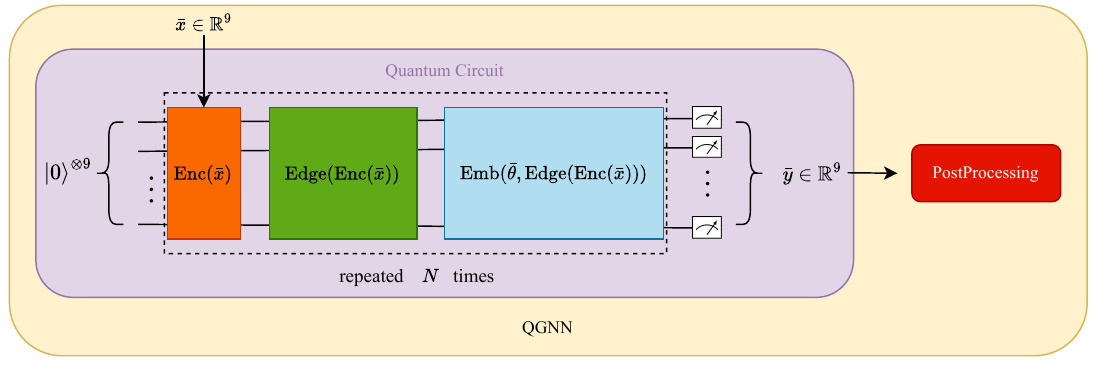}
    \caption{The QGNN schema.}
    \label{scheme}
\end{figure}

The QGNN architecture proposed here represents a fusion of classical GNNs with quantum computing principles. 
Here we introduce the architecture of the proposed QGNN model, depicted in Fig.~\ref{scheme}.

The elements of the architecture are the following ones:
\begin{enumerate}
    \item \textbf{Data encoding.} In this step, information about the datapoint (i.e., the coordinates of the atoms composing the molecule) is encoded in the quantum circuit.
    \item \textbf{Edge layer.} In the edge layer, information regarding the distances between each pair of atoms is passed.
    \item \textbf{Learnable embedding layer.} This is the main block of the architecture, which learns the embeddings of the nodes that are constantly updated to later measure them for predicting the forces.
    \item \textbf{Repeat and reupload.} The previous steps are repeated $N$ times, where $N$ is the number of layers desired for the architecture. At the beginning of each layer, data-reuploading \cite{reuploading} is addressed by applying the encoding layer once again.
    \item \textbf{Measure of the qubits.} All the qubits are measured to obtain the unprocessed force values.
    \item \textbf{Post-processing (PP) and prediction.} A learnable transformation is applied to the output of the circuit, and the prediction for the forces is made, from which the energy is predicted by pooling.
\end{enumerate}

In the following each step of the QGNN is described in detail. A complete layer is defined as the sequence of encoding, edge, and embedding layers. The Appendix~\ref{appendix} contains the definitions of the gates used in these layers.

\subsubsection{Encoding layer}
\begin{figure}[!ht]
\centering
    \includegraphics[scale=.85]{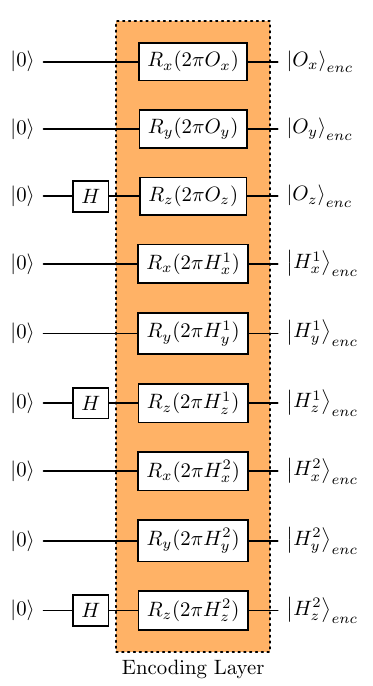}   
    \caption{Encoding layer. As this figure visualizes the first layer of the architecture, the qubits are initialized in $\ket{0}$.}
    \label{enc}
\end{figure}

The architecture employed is based on the nature of the problem addressed. 9 qubits are initialized in the $\ket{0}$ state, with each qubit representing an input coordinate, which allows, with an excess of notation, to refer to the qubit's state as $\ket{\alpha_c}$.
For $z$-coordinate qubits a Hadamard gate is applied in order to get an initial superposition of states, which allows to have initialization of the parameters to zero without compromising the way in which the architecture works. \newline
It's important to recall that the coordinates have been rescaled to the $[0,1]$ range, it allows to create an intuitive encoding by performing rotations on each qubit along the axis it represents, as follows:
\begin{equation}
    \forall \alpha \in \mathcal{A},\, c \in \mathcal{C}, \quad
\text{Enc}(\ket{\alpha_c}) := R_c(2 \pi \alpha_c) \ket{\alpha_c} = \ket{\alpha_c}_{\text{enc}}.
\end{equation}
The encoding layer circuit is reported in Fig.~\ref{enc}.

\subsubsection{Edge layer}\label{edge_l}
\begin{figure}[!ht]
\centering
    \includegraphics[width=\textwidth]{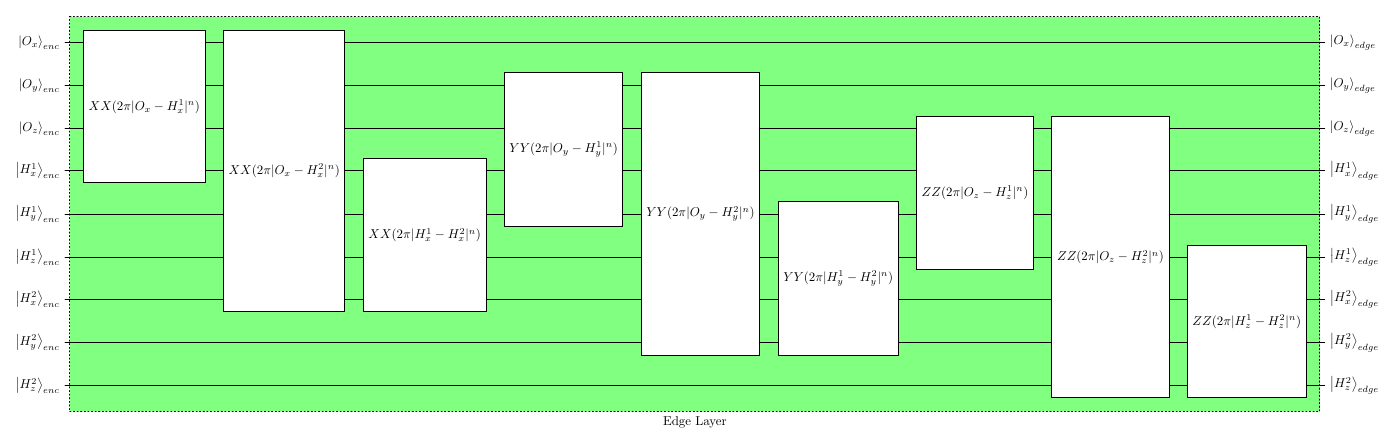}   
    \caption{Edge layer of the $n$-th complete layer.}
    \label{edge}
\end{figure}

The aim of the edge layer is to encode in the circuit the information about the distances between the atoms in the molecule. In order to do so, $XX,$ $YY$, and $ZZ$ rotation gates are employed. $XX$ gates are used to entangle $x$-coordinate qubits, $YY$ gates for $y$-coordinate qubits and $ZZ$ gates for $z$-coordinate qubits. The angle $\theta$ by which the single qubit is rotated is proportional to the distance between the atoms along the coordinate, which allows to define the operation of the edge layer on a single qubit as:

\begin{equation}
\begin{aligned}
\forall \alpha \in \mathcal{A},\, c \in \mathcal{C}, \quad
\text{ Edge}\left(\ket{\alpha_c}_{\text{enc}}\right) & := \prod_{j \in \mathcal{A}} CC\left(2\pi \abs{\alpha_c - j_c}^n\right)\ket{\alpha_c}_{\text{enc}} = \ket{\alpha_c}_{\text{edge}}, \\
\mathcal{A} & = 
\begin{cases} 
\left\{H^1, H^2\right\} & \text{if } \alpha = O, \\
\left\{H^2\right\} & \text{if } \alpha = H^1, H^2
\end{cases}
\end{aligned}
\end{equation}
where $CC \in \left\{XX, YY, ZZ \right\}$ is the gate associated with the coordinate $c \in \mathcal{C}$,$|\alpha_c - j_c|$ comes from the previously defined $D$ tensor, whose entries are rescaled in $[0,1]$, and $n={1,2...,N}$, corresponds the complete layer currently in use. Raising the distance to the power of $n$ allows the model to grasp different order interactions between the atoms, thus enhancing the expressiveness of the architecture.
The edge layer circuit is reported in Fig.~\ref{edge}.

\subsubsection{Learnable embedding layer}
This is the core of the architecture, where the node embeddings are updated. For each qubit, a rotation by a learnable parameter $\theta_k$ is performed along the axes not represented by that qubit, resulting in 18 learnable parameters per layer. Additionally, if the current layer is not the last one, SWAP gates are added between the $x$, $y$, and $z$ coordinates of the two hydrogen atoms. SWAP gates are used to exchange the states of two qubits and combined with parameter updates through backpropagation, allowing simultaneous and interdependent weight updates of the qubits involved in the SWAP operation. This induces a weight-induced permutation invariance, which is a property of the molecule being modeled and is reflected in the model architecture.

Let's introduce the learnable operation of the embedding layer $\lambda$ to then define the layer action:
\begin{equation}
    \lambda \ket{\alpha_c} = \prod_{i\in \mathcal{C}^-(\alpha_c)} R_i \left(2\pi \theta_{k,n}\right)\ket{\alpha_c},\quad \mathcal{C}^-(\alpha_c) = \{x,y,z\} \setminus \{c\},
\end{equation}
where $k=1\dots 18$ refers to the index of the learnable parameter and $n$ to the complete layer. The action of the embedding layer on the single qubit can then be defined as:

\begin{equation}
\begin{aligned}
\forall \alpha \in \mathcal{A},\, c \in \mathcal{C}, \quad
\text{Emb}\left(\ket{\alpha_c}_{\text{edge}}\right) = \ket{\alpha_c}_{\text{emb}} =
\begin{cases}
\lambda \ket{\alpha_c}_{\text{edge}} & \text{if } \alpha \in \{O,H^2\}, \\
\text{SWAP}\left(\lambda \ket{\alpha_c}_{\text{edge}}, \lambda \ket{H^2_c}_{\text{edge}}\right)  & \text{if } \alpha = H^1.
\end{cases}
\end{aligned}
\end{equation}

\begin{figure}[!ht]
\centering
    \includegraphics[scale=.85]{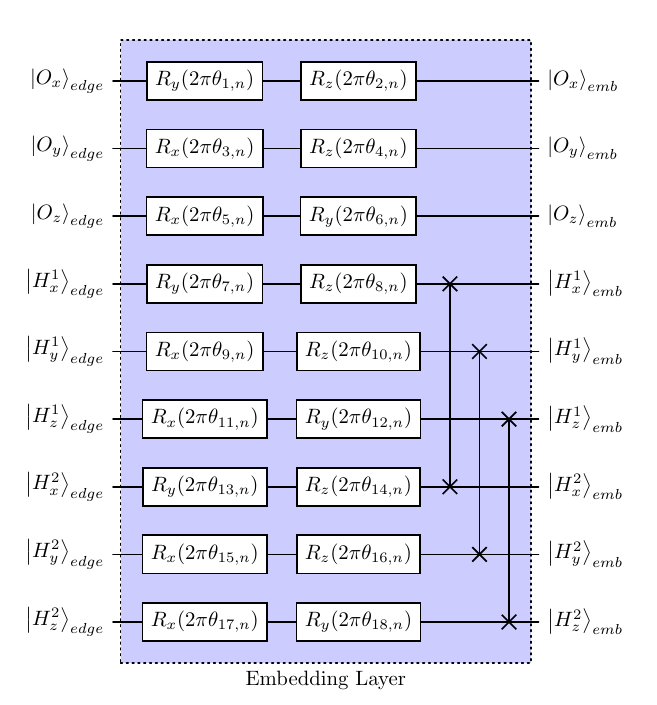}
    \caption{Embedding layer of the $n$-th complete layer.}
    \label{emb}
\end{figure}

The embedding layer circuit is reported in Fig.~\ref{emb}.

\subsubsection{Layer reuploading}
A complete layer of the architecture is defined as the sequence of one encoding layer, one edge layer, and one embedding layer. Adding subsequent complete layers increases the expressibility of the architecture. An analysis of the expressibility in relation to the number of layers used is provided in Sect. \ref{RES}.

The application of the data reuploading technique simply means that the encoding layer is reapplied to the considered layer stack. Additionally, the behavior of the edge layer changes from layer to layer: the power to which the distance between the atoms' coordinates is raised is adjusted to emphasize different order interactions.

For completeness and to simplify the understanding of the architecture, Fig.~\ref{2layer} illustrates what a 2-layer QGNN circuit looks like:

\begin{figure}[!ht]
  \centering
    \includegraphics[width=0.9\textwidth]{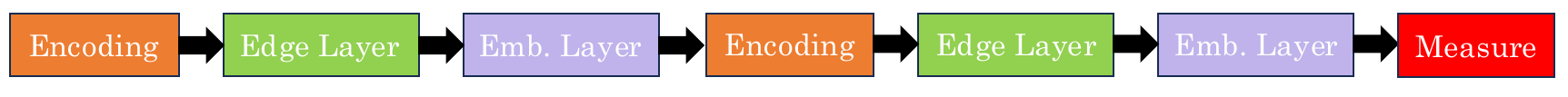}   
    \caption{Unrolled representation of a QGNN circuit with 2 layers.}
    \label{2layer}
\end{figure}

\subsubsection{Measure of the qubits}
Each qubit is measured individually using the expectation value of the Pauli operator $\sigma_c$ applied to the qubit's state in output from the last layer $\ket{\alpha_c}_{\text{emb}}$. 
The expectation value of $\sigma_c$ is determined by applying the corresponding measurement operator to the qubit and collecting measurement outcomes. This process enables us to obtain the unprocessed value of the forces along each axis, which means that if we measure the qubit in state $\ket{\alpha_c}_{\text{emb}}$, we will obtain the unprocessed resulting force $F^{\alpha_c}_u$.
\subsubsection{Post-processing and prediction}
\begin{figure}[!ht]
    \centering
    \includegraphics[width=0.6\textwidth]{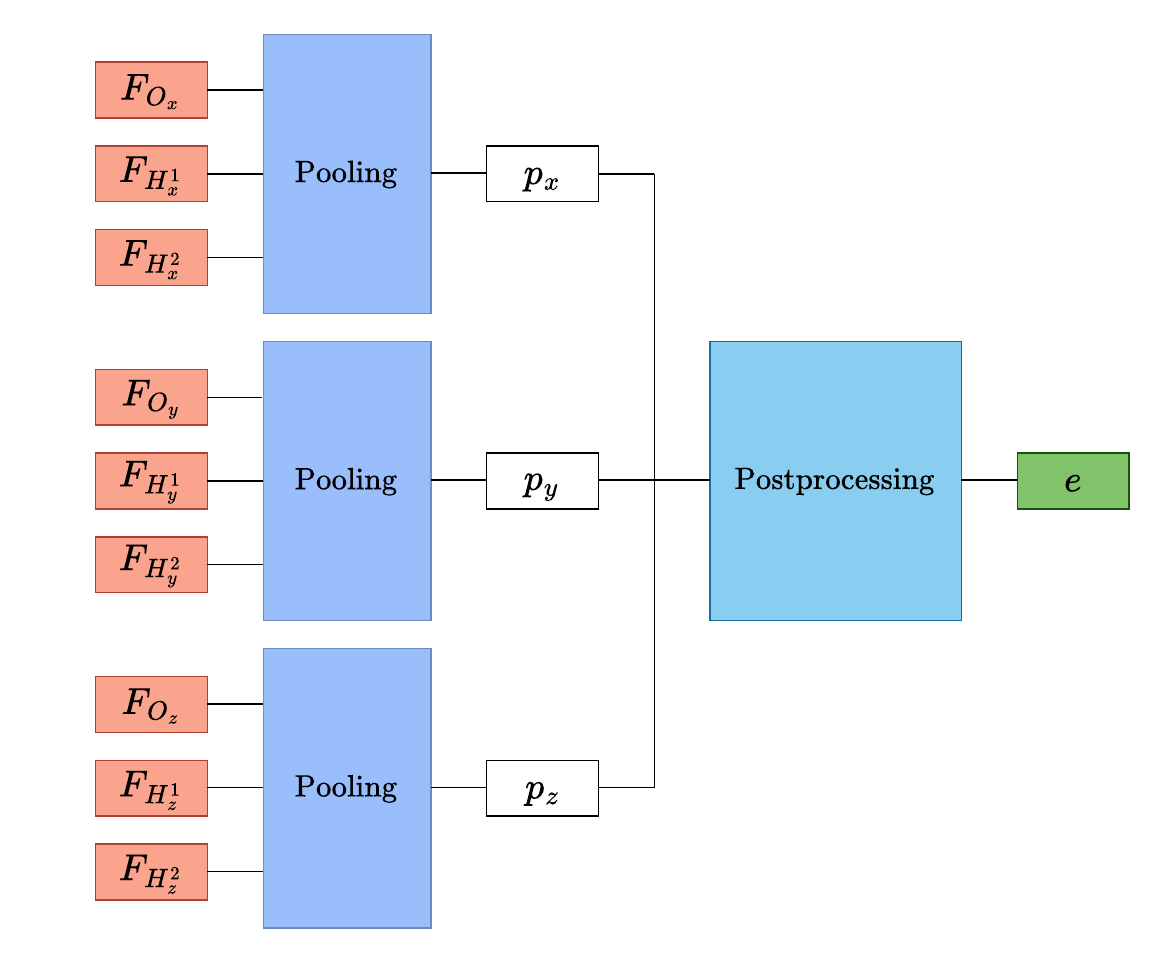}   
    \caption{Pooling for energy prediction.}
    \label{pool}
\end{figure}

Following the measurement of qubits, 9 raw force values are obtained, each within the range $[-1,1]$. To derive the precise force values from this set, a custom Fully Connected layer with a reduced number of parameters is employed, facilitating efficient processing and extraction of the desired force data.
In practice, each raw value is multiplied by a learnable parameter $k_{\alpha_c}$ and then added to a common learnable bias $b$, which is consistent across all 9 values. The resulting value represents the force projection as follows
\begin{equation}
\forall \alpha \in \mathcal{A},\, c \in \mathcal{C}, \quad F^{\alpha_c} = k_{\alpha_c} F^{\alpha_c}_u + b.
\end{equation}
The obtained 9 force values are subsequently used to predict the energy, which is a global feature of the graph. Pooling is employed to extract this global feature, as illustrated in Fig.~\ref{pool}.

To perform Sum Pooling \cite{BabenkoL15}, the values of the forces extracted along each axis are summed, resulting in three values, one per coordinate, denoted as $p_x$, $p_y$, and $p_z$. Similarly to post-processing of the forces, a shared bias is used for these three values, with one learnable parameter assigned to each, which are used to predict the total energy $e$ of the molecule.
It should be noted that, for improved performance, a different approach is taken during training and testing: the values used for pooling during training are the actual forces, whereas during testing, the predicted forces are utilized.

Two variations of the described model were tested. The first variation consists of 2 layers with a total of 50 parameters, of which 36 are from the quantum circuit (18 parameters per layer) and 14 are from the post-processing custom Fully Connected layers. The second variation features the same architecture but with 3 layers, totaling 68 parameters, with 54 parameters from the quantum circuit and 14 from the classical post-processing.
Models' parameters are initialized as follows:
\begin{itemize}
    \item \textbf{Quantum circuit parameters} have been randomly initialized from a normal distribution $\mathcal{N}(\mu = 0, \, \sigma=1)$;
    \item \textbf{Post-Processing traslational parameters} have been randomly initialized from a uniform distribution $\mathcal{U}(-0.2, \,  0.2)$;
    \item \textbf{Post-Processing scale parameters} have been randomly initialized from a uniform distribution $\mathcal{U}(1, \,  1.5)$.
\end{itemize}

\section{Experimental Assessment}
\label{chap: Chapter 5}

\subsection{Experimental Setting}
The loss function used is a combination of Mean Squared Error (MSE) and a KL-Divergence inspired term (KLI). By minimizing the MSE, the objective is to ensure that the predictions are as close as possible to the true values, while KLI is employed to make the output distribution as close as possible to the true distribution. We can't use the exact KL-Divergence term because we are not dealing with probability distributions.
Thus, for each batch, the loss is computed as:
\begin{equation}
    \mathcal{L}^{\text{batch}} = \text{MSE}_{\text{batch}}  + \gamma \, \text{KLI}_{\text{batch}}^2 
\end{equation}
where $\gamma$ is a hyperparameter used to weight the KLI contribution. 

The MSE of a batch is computed as:
\begin{equation}
\text{MSE}_{\text{batch}} = \frac{1}{b_s} \sum_i \Biggl(  (e_i - \hat{e}_i)^2  + \sum_{\alpha \in \{O,H^1,H^2\}} \, \sum_{c \in \{x,y,z\}} (F_i^{\alpha_c} - \hat{F}_i^{\alpha_c})^2 \Biggr)
\end{equation}
where $b_s$ represents the batch size, the index $i$ runs on the elements of the batch, $e_i$ and $F_i^{\alpha_c}$ represent the true values of the energies and the resulting forces on each atom along each axis $c$, respectively, while $\hat{e}_i$ and $\hat{F}_i^{\alpha_c}$ are respectively the predicted values of the energies and the resulting forces on each atom along each axis $c$.

The second contribution of the loss, i.e. the squared KLI of a batch, is computed as:
\begin{equation}
\text{KLI}_{\text{batch}}^2 = \frac{1}{b_s} \sum_i \left( |e_i|^2     \log \Bigg| \frac{e_i}{\hat{e}_i + \varepsilon} \Bigg|^2 + 
\sum_{\alpha \in \{O,H^1,H^2\}} \, \sum_{c \in \{x,y,z\}} |F_i^{\alpha_c}|^2       \log \Bigg| \frac{F_i^{\alpha_c}}{\hat{F}_i^{\alpha_c} + \varepsilon} \Bigg|^2 \right) 
\end{equation}
where $\varepsilon$ small, of the order of $10^{-8}$, is needed not to occur in computation errors.

To compute the total loss \(\mathcal{L}\), the average over the number of batches used is calculated, resulting in:
\begin{equation}
\mathcal{L} = \frac{1}{n_b} \sum_{i=1}^{n_b} \mathcal{L}^{\text{batch}}_i
\end{equation}
where $n_b$ is the number of batches we are making use of and $\mathcal{L}^{\text{batch}}_i$ represents the loss computed on the $i$-th batch.

The evaluation metric employed to assess the results is the Root Mean Square Error (RMSE). This choice was made to ensure consistency with similar works, such as \cite{oriel}.

The model training was conducted using four Tesla V100S-PCIE-32GB GPUs to ensure efficient simulation of the 9-qubit circuit while maintaining high computational speed. The QGNN was implemented in Python 3.11.4 with the assistance of Pennylane 0.30.0 for simulating quantum circuits. The training process was facilitated using Jax 0.3.25 for fast quantum circuit simulation on a noiseless backend. The Adam optimizer \cite{adam} with a learning rate of 0.001 was employed while incorporating an early stopping mechanism to optimize training efficiency. Further details regarding the hyperparameters, which have been obtained through an extensive grid search procedure, can be found in Table~\ref{tab:hyperparameters} for reference.
\begin{table}[!ht]
\centering
\caption{Hyperparameters Used for Training the Adopted Model}
\renewcommand{\arraystretch}{1.4}
\begin{tabular}{lr}
\toprule
Hyperparameter & Value \\
\midrule
$b_s$ & 128 \\
$n_b$ & 69 \\
$\gamma$ & 0.03125 \\
Number of epochs & 500 \\
Patience (early stopping) & 50 \\
\bottomrule
\end{tabular}
\label{tab:hyperparameters}
\end{table}
It is noted that the parameters used at test time are those that retain the lowest validation loss obtained during training.

\subsection{Numerical Results}
\label{RES}
\begin{figure}[!ht]
    \centering
    \includegraphics[width=10cm]{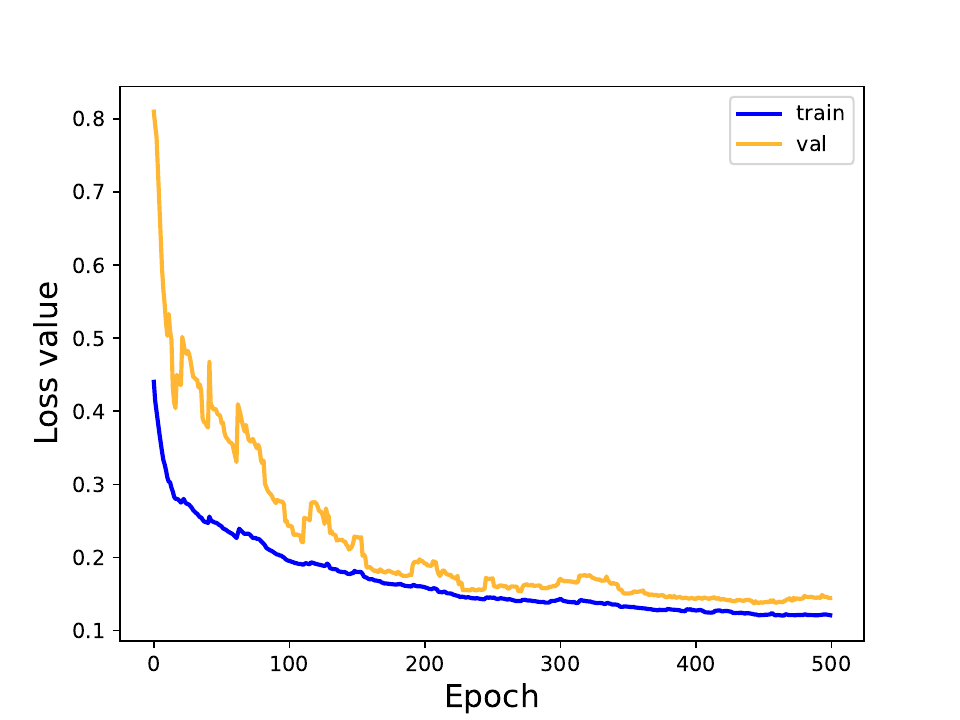}   
    \caption{Train and validation losses for our 2-layer QGNN architecture.}
    \label{fig:loss1}
\end{figure}

In this section, the model's performances during the training and test phase are discussed. The train and validation losses are illustrated in Fig.~\ref{fig:loss1}. Both losses are decreasing in a very smooth way, which means that the model is learning without overfitting, which is due to good tuning of the parameters.

\begin{figure}[!ht]
    \centering
    \includegraphics[width=10cm]{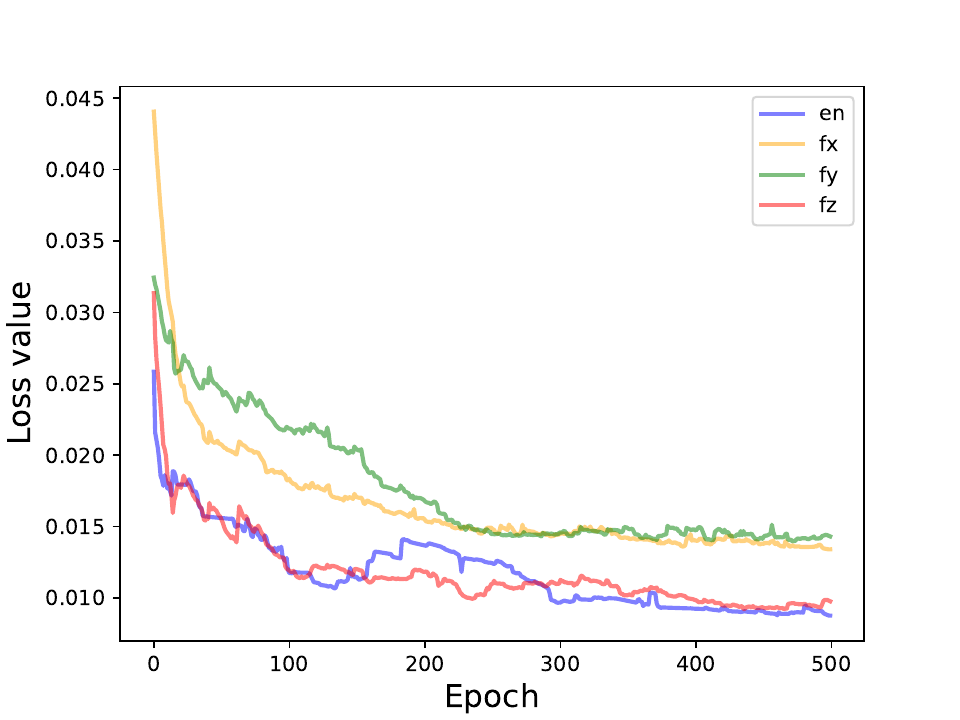}  
    \caption{Components of the train loss for our 2-layer QGNN architecture.}
    \label{fig:loss2}
\end{figure}

The train loss is further analyzed by examining its components, where the $x$, $y$, and $z$ components are summed separately, as shown in Fig.~\ref{fig:loss2}. This analysis was conducted to determine if any component is predominant in contributing to the total loss. It is observed that while the $z$ component is slightly lower than the others, their magnitudes do not differ significantly.

The performances obtained by the model on the validation and test set are reported in Table~\ref{tab: performances}, compared to other state-of-the-art methods.
\begin{table}[!ht] 
\centering
\footnotesize
\caption{Results Comparison} 
\begin{tabular}{lccccc}
\toprule
Model & $\#$params & RMSE(E) & RMSE(F)& RMSE(E)& RMSE(F)\\
 & & [val] & [val]& [test] & [test]\\
\midrule
2-layer QGNN & 50 & 0.0874& 0.0731& 0.0844 &0.0694\\
3-layer QGNN & 68 & 0.0963& 0.0696& 0.0916&0.0675\\
QNN \cite{oriel} & 87 & 0.0050& 0.0600& \textbackslash & \textbackslash \\
\bottomrule
\end{tabular}
\label{tab: performances}
\end{table}
Not all the necessary details relevant to data points used for training and testing were mentioned in \cite{oriel}. So, herein the comparison is made considering the respective (optimal) setups chosen in that paper and in the present one, although the hyperparameters values are not necessarily identical. 

In molecular interactions, electromagnetic forces are expected to have the most significant impact on atomic interactions. These forces exhibit a scaling behavior proportional to $r^2$, where $r$ denotes the distance between the interacting atoms. It is noteworthy that the architecture effectively captures this behavior through a customized use of the edge layer, which highlights different order interactions. Additionally, the architecture produces highly similar results for both 2 and 3 layers of the QGNN. This indicates that the architecture can accurately reconstruct and replicate the underlying physics of the dataset, even with a low circuit depth, which is essential for practical implementation on NISQ devices.

\begin{figure}[!ht]
    \centering
    \includegraphics[width=.55\textwidth]{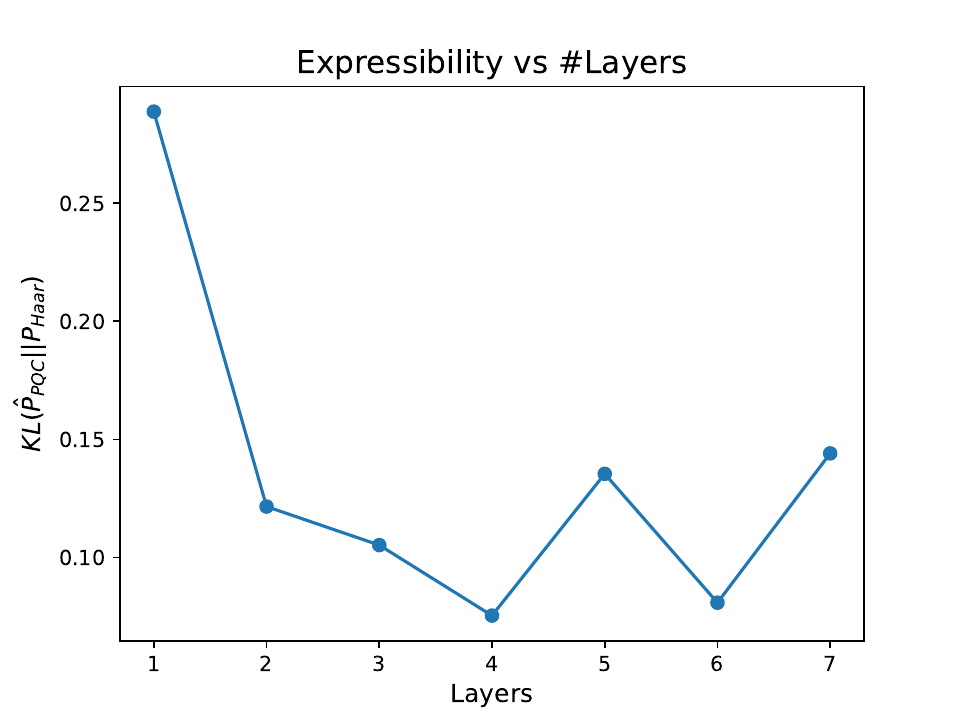}   
    \caption{Expressibility of the QGNN by the number of layers.}
    \label{expr}
\end{figure}

The expressibility of the QGNN is evaluated based on the number of layers employed. For the definition of expressibility, refer to \cite{expr}. Expressibility measures the circuit’s ability to produce quantum states that are representative of the Hilbert space. To quantify this feature of the VQC, the distribution of states obtained from the circuit with randomly sampled parameters is compared to the uniform distribution of states, i.e., the ensemble of Haar-random states. In practical terms, this can be computed as the KL divergence between the estimated fidelity distribution $\hat{P}_{PQC}$ produced by a PQC and that of the Haar-distributed ensemble $P_{Haar}$:
\begin{equation}
\mathcal{E} = D_{KL}(\hat{P}_{PQC}||P_{Haar}).
\end{equation}
A PQC yielding a fidelity distribution with a lower KL divergence compared to the Haar distribution indicates a circuit with higher expressibility. Consequently, we investigate how expressibility changes with the number of layers in the novel architecture, as illustrated in Fig.~\ref{expr}.

A clear trend emerges from Fig.~\ref{expr}: with the progressive addition of layers, the circuit's expressibility also grows, as highlighted by the decreasing KL divergence. This pattern holds for the initial four layers; however, beyond this point, oscillations in the function become apparent. This aligns with our anticipations, suggesting an eventual plateau where expressibility reaches its maximum. Additional experiments with more layers may illustrate the function consistently oscillating around this plateau value.


While the primary focus of this work revolved around the introduction of a novel architecture, it is worth comparing the results with those obtained in \cite{oriel}. Notably, the comparison reveals that similar performance levels in force predictions were achieved using approximately $60\%$ of the parameters employed in \cite{oriel}. This achievement is particularly significant since our approach substantially differs from the previous works.
In\cite{oriel}, the methodology primarily focused on predicting the energy and analytically deriving force components through differentiation using the parameter shift rule \cite{qcl}. In contrast, our approach tackles a more complex task by aiming to predict all ten underlying target variables, i.e. nine intramolecular force components and the total energy, instead of just predicting the energy alone. Consequently, obtaining comparable results with a reduced number of parameters is a solid achievement.

Moreover, the fact that we achieve lower loss on the test set compared to the validation set is an encouraging result: it suggests that our model can generalize effectively based on the training data, which aligns with the use of early stopping on the validation set during training.
It is anticipated to obtain a higher RMSE for the energy predictions compared to \cite{oriel}. This problem arises from our approach of post-processing the 9 forces to extract the energy, effectively propagating the errors from force predictions. As we reflect on these results, it becomes evident that future research should explore alternative methods for inferring the energy. One potential avenue for improvement involves circuit modification, such as introducing a 10-th qubit dedicated to energy measurements. However, this approach warrants further exploration and thoughtful consideration to ensure its effectiveness.


\section{Conclusions}
\label{chap: Chapter 6} 


In this paper, a novel architecture based on QGNNs applied to molecular physics is introduced. It takes a distinct approach compared to previous attempts at implementing QGNNs and demonstrates the feasibility of obtaining node embeddings and updates in a manner conceptually akin to classical GNNs. This introducotry work has provided highly satisfying outcomes.

One of the key strengths of the architecture lies in its adeptness at harnessing the problem's inherent symmetry, which is achieved by utilizing SWAP gates between the two hydrogen atoms, effectively enforcing a form of permutation invariance driven by the weights. To the best of our knowledge, this approach has not been explored previously and significantly streamlines the learning process.
Another innovative aspect of the QGNN is in the utilization of edge layers. As extensively discussed in Sect.~\ref{RES}, the strategy of passing different-order distances to the layers has proven to be highly effective: it equips the architecture with the ability to capture the underlying physics of the target process.


\section*{Acknowledgment}
The contribution of A. Ceschini and M. Panella in this work was supported by the ``NATIONAL CENTRE FOR HPC, BIG DATA AND QUANTUM COMPUTING'' (CN1, Spoke 10) within the Italian ``Piano Nazionale di Ripresa e Resilienza (PNRR)'', Mission 4 Component 2 Investment 1.4 funded by the European Union - NextGenerationEU - CN00000013 - CUP B83C22002940006.

\bibliographystyle{IEEEbib}
\bibliography{bib}

\appendix
\label{appendix}
%
We present in the following the gates adopted to realize the proposed QGNN circuit.

\noindent\textbf{$\sigma_x$ gate.}\\
In quantum computing the NOT operation is performed by the Pauli-X gate.
The matrix that defines the $\sigma_x$ gate is the following:
\begin{center}
$\sigma_x = \begin{pmatrix}0 & 1\\1 & 0\end{pmatrix}$
\end{center}

\noindent\textbf{$\sigma_y$ gate.}\\
The $\sigma_y$ gate is defined as
$$\sigma_y = \begin{pmatrix}0 & -i\\i & 0\end{pmatrix}$$

\noindent\textbf{$\sigma_z$ gate.}\\
As for the previous ones we have the $\sigma_z$ gate, which is defined as 
$$\sigma_z = \begin{pmatrix}1 & 0\\0 & -1\end{pmatrix}$$

\noindent\textbf{H gate.}\\
Another fundamental gate for quantum computing is the Hadamard gate, or H gate. 
The matrix that defines the H gate is the following:
\begin{center}
$H = \cfrac{1}{\sqrt{2}}\begin{pmatrix}1 & 1\\1 & -1\end{pmatrix}$, 
\end{center}
This gate prepares an equal superposition of all the computational basis in the Hilbert space so that, if the state is measured, there is an equal chance to get $\ket{0}$ or $\ket{1}$ as a result. From a circuital point of view, it is denoted by the symbol:
\begin{center}
\begin{quantikz}
\lstick{}\qw & \qw & \gate{H} & \rstick{}\qw & \qw\\
\end{quantikz}
\end{center}

\noindent\textbf{$R_x$ gate.}\\
The $R_x$ gate is one of the rotation operators. It is a single-qubit rotation of an angle $\theta$ around the $x$-axis and is defined as:
$$R_x(\theta) = \begin{pmatrix} \cos(\frac{\theta}{2}) & -i\sin(\frac{\theta}{2})\\-i\sin(\frac{\theta}{2}) & \cos(\frac{\theta}{2})\end{pmatrix}$$
and is represented as:
\begin{center}
\begin{quantikz}
\lstick{}\qw & \qw & \gate{R_x} & \rstick{}\qw & \qw\\
\end{quantikz}
\end{center}

\noindent\textbf{$R_y$ gate.}\\
Analogously to the $R_x$ gate, the $R_y$ gate is a single-qubit rotation of an angle $\theta$ around the $y$-axis and is defined as:
$$R_y(\theta) = \begin{pmatrix} \cos(\frac{\theta}{2}) & -\sin(\frac{\theta}{2})\\\sin(\frac{\theta}{2}) & \cos(\frac{\theta}{2})\end{pmatrix}$$
and is represented as:
\begin{center}
\begin{quantikz}
\lstick{}\qw & \qw & \gate{R_y} & \rstick{}\qw & \qw\\
\end{quantikz}
\end{center}

\noindent\textbf{$R_z$ gate.}\\
We also have a $z$-axis rotation which is defined as:
$$R_z(\theta) = \begin{pmatrix} e^{-i \frac{\theta}{2}} & 0\\0& e^{i \frac{\theta}{2}}\end{pmatrix}$$
and is represented as:
\begin{center}
\begin{quantikz}
\lstick{}\qw & \qw & \gate{R_z} & \rstick{}\qw & \qw\\
\end{quantikz}
\end{center}

\noindent\textbf{SWAP gate.}\label{swapp}\\
The SWAP gate is a 2-qubit gate which simply exchanges the bit values it is passed and is defined as follows:
$$\text{SWAP} = \begin{pmatrix}1 & 0 & 0 & 0\\
 0 & 0 & 1 & 0\\
 0 & 1 & 0 & 0\\
 0 & 0 & 0 & 1\\ \end{pmatrix}$$
It is represented as:
\begin{center}
\begin{quantikz}
& \swap{1} & \qw \\
& \targX{} & \qw
\end{quantikz}
\end{center}

\noindent\textbf{$XX$ Ising gate.}\\
The $XX(\theta)$ Ising gate is a parametric 2-qubit interaction gate. It's symmetric and maximally entangling at $\theta = \frac{\pi}{2}$. It is defined as:
$$XX(\theta) = e^{-i \frac{\theta}{2} (\sigma_x \otimes \sigma_x)} = 
\begin{pmatrix} \cos(\frac{\theta}{2}) & 0 & 0 &-i\sin(\frac{\theta}{2})\\
 0 & \cos(\frac{\theta}{2}) & -i\sin(\frac{\theta}{2}) & 0\\
 0 & -i\sin(\frac{\theta}{2}) & \cos(\frac{\theta}{2}) & 0\\
 -i\sin(\frac{\theta}{2}) & 0 & 0 &\cos(\frac{\theta}{2})\\ \end{pmatrix}$$
and is represented as:
\begin{center}
    \begin{quantikz}
        &\gate[2]{XX}& \qw \\
        &  & \qw
    \end{quantikz}
\end{center}

\noindent\textbf{$YY$ Ising gate.}\\
The $YY(\theta)$ Ising gate is a parametric 2-qubit interaction gate. It is analogous to the $XX$ gate and is defined as:
$$YY(\theta) = e^{-i \frac{\theta}{2} (\sigma_y \otimes \sigma_y)} = 
\begin{pmatrix} \cos(\frac{\theta}{2}) & 0 & 0 &i\sin(\frac{\theta}{2})\\
 0 & \cos(\frac{\theta}{2}) & -i\sin(\frac{\theta}{2}) & 0\\
 0 & -i\sin(\frac{\theta}{2}) & \cos(\frac{\theta}{2}) & 0\\
 i\sin(\frac{\theta}{2}) & 0 & 0 &\cos(\frac{\theta}{2})\\ \end{pmatrix}$$
and is represented as:
\begin{center}
\begin{quantikz}
& \gate[2]{YY}& \qw \\
&  & \qw
\end{quantikz}
\end{center}

\noindent\textbf{$ZZ$ Ising gate.}\\
The $ZZ(\theta)$ Ising gate is a parametric 2-qubit interaction gate. It's analogous to the gates we just saw and is defined as:
$$ZZ(\theta) = e^{-i \frac{\theta}{2} (\sigma_z \otimes \sigma_z)} = 
\begin{pmatrix} e^{-i \frac{\theta}{2}} & 0 & 0 & 0\\
 0 & e^{i \frac{\theta}{2}} & 0 & 0\\
 0 & 0 & e^{i \frac{\theta}{2}} & 0\\
 0 & 0 & 0 & e^{-i \frac{\theta}{2}}\\ \end{pmatrix}$$
and is represented as:
\begin{center}
\begin{quantikz}
& \gate[2]{ZZ}& \qw \\
&  & \qw
\end{quantikz}
\end{center}

\end{document}